\documentclass[prl,aps,twocolumn,showpacs,floatfix]{revtex4-1}
\usepackage{amsfonts}
\usepackage{amssymb}
\usepackage{hyperref}
\usepackage{graphicx}
\usepackage{dcolumn}
\usepackage{bm,amsmath,verbatim}
\usepackage{mathrsfs}
\usepackage{color}
\usepackage{times}
\usepackage{bm}

\hypersetup{colorlinks,
linkcolor=blue,          citecolor=blue,        filecolor=blue,      urlcolor=blue           }

\def\be{\begin{equation}}
\def\ee{\end{equation}}
\def\bea{\begin{eqnarray}}
\def\eea{\end{eqnarray}}
\def\bse{\begin{subequations}}
\def\ese{\end{subequations}}

\def\be{\begin{eqnarray}}
\def\ee{\end{eqnarray}}

\begin{document}

\title{Observation of Dynamical Quantum Phase Transition
with Correspondence \\in Excited State Phase Diagram}

\author{T. Tian}
\thanks{These authors contributed equally to this work.}
\author{H.-X. Yang}
\thanks{These authors contributed equally to this work.}
\author{L.-Y. Qiu}
\author{H.-Y. Liang}
\author{Y.-B. Yang}
\author{Y. Xu}
\email{yongxuphy@tsinghua.edu.cn}
\author{L.-M. Duan}
\email{lmduan@tsinghua.edu.cn}
\affiliation{Center for Quantum Information, IIIS, Tsinghua University, Beijing 100084,
PR China}

\begin{abstract}
Dynamical quantum phase transitions are closely related to equilibrium quantum phase transitions for
ground states. Here, we report an experimental observation of a dynamical quantum phase transition
in a spinor condensate with correspondence in an excited state phase diagram, instead of the ground state one.
We observe that the quench dynamics exhibits a non-analytical change with respect to a parameter in the final Hamiltonian
in the absence of a corresponding phase transition for the ground state there. We make a connection between this singular point
and a phase transition point for the highest energy level in a subspace with zero spin magnetization
of a Hamiltonian. We further show the existence of dynamical phase transitions for finite magnetization
corresponding to the phase transition of the highest energy level in the subspace with the same magnetization.
Our results open a door for using dynamical phase transitions
as a tool to probe physics at higher energy eigenlevels of many-body Hamiltonians.
\end{abstract}

\maketitle

Non-equilibrium quantum many-body dynamics have seen a rapid progress in recent years
due to deepened theoretical understanding~\cite{Vengalattore2011RMP,Heyl2018RPP,Zvyagin2016LTP,Fabrizio2016PTRA} and experimental technology advances in systems, such as
trapped ions~\cite{Monroe2017Nat,Roos2017PRL}, Rydberg atoms~\cite{Lukin2017Nat}, ultracold atoms~\cite{Weitenberg2018NP,ShuaiChen2018PRL,Yang2019PRA,SmaleSciAdv2019},
nitrogen-vacancy centers~\cite{Lukin2017Nat2}, and others~\cite{GuoPRAP2019}. One central question in the field concerns the
existence of phase transitions as a system parameter is suddenly varied (referred to as dynamical
quantum phase transitions~\cite{Heyl2018RPP,Zvyagin2016LTP,Fabrizio2016PTRA})). Based on different identification features, such a phase transition
can generally be divided into two types. One type refers to the existence of a non-analytical behavior
in a long time steady state of a local order parameter with respect to a final Hamiltonian parameter~\cite{Altshuler2006PRL,Biroli2010PRL}.
The other type corresponds to the emergence of a singularity in a global order parameter such as
Loschmidt echoes with respect to time after a quench~\cite{Heyl2013PRL,Silva2018PRL}. Both of these two types of dynamical phase transitions are closely related to the ground state quantum phase transition.
However,
exceptions exist and the Loschmidt echo
is allowed to show non-analytical behavior even though a system parameter is quenched within an
identical ground state phase~\cite{Andraschko2014PRB,Fagotti2013,Vajna2014PRB,Halimeh2017PRB,Weitenberg2018NP}. Moreover,
whether the dynamical phase transition with no correspondence in ground state phase diagram
is related to an excited state quantum phase transition is
still an open question~\cite{Heyl2018RPP,Cejnar2006JPA,Caprio2008AP,Cejnar2011PRA,Wambach2013PRB,Santos2015PRA}.

Similar to the ground state quantum phase transitions, excited state quantum phase transitions
refer to the existence of singularities in the energy or an order parameter of an excited energy level~\cite{Cejnar2006JPA,Caprio2008AP}.
While such a phase transition has been proposed for more than a decade,
it has not been experimentally observed in a many-body quantum system.
Recently, Ref.~\cite{12} has theoretically
proposed a dynamical phase transition that is closely related to the quantum phase transition
for the highest energy level in a subspace with zero spin magnetization in a spinor condensate.
From this perspective, the spinor condensate provides an ideal experimental many-body quantum platform
for probing the excited
state quantum phase transitions by quench dynamics. In fact, many non-equilibrium phenomena, such
as spin domains, topological defects and Kibble-Zurek mechanism, have been experimentally
observed in a spinor condensate~\cite{StamperKurn2006Nat,Raman2011PRL,Vinit2013PRL,HoangNatComm2016,Anquez2016PRL,KimPRL2017,
PruferNature2018,Shin2019PRL,Gerbier2019NC,Chen2019PRL,Shin2019arXiv}.
In addition, the highest energy level in the subspace has an upper bound in energy in a finite
system, reminiscent of a state with a negative absolute
zero temperature, which has been experimentally realized~\cite{Pound1951,Lounasmaa1997,Ketterle2011,Lounasmaa1994,Bloch2013}.

In this paper, we report the experimental observation of a dynamical quantum phase transition
with correspondence in the highest energy level phase diagram in a subspace with fixed spin magnetization
in a spinor condensate.
Instead of measuring a long time
steady value of an order parameter such as the number of atoms with zero spin, we probe the value of the first peak
of the time evolution of the atom number appearing in a short time.
By preparing a condensate in an antiferromagnetic (AFM) state,
we find that the quench dynamics show a non-analytical change
as a function of the quadratic Zeeman energy of a final Hamiltonian at $q_f=2c_2$ ($c_2$ describes an interaction strength) as
$q$ is suddenly varied from a large negative value to $q_f$. Our results are beyond the ground state phase transition given the absence of a phase transition at $q=2c_2$. However, our finding is highly related to the phase transition
between an AFM and a broken-axisymmetry (BA) phase for the highest energy level in the subspace with zero spin magnetization. We further measure
the quench dynamics for finite magnetization and find singular behaviors determined by the phase transition
on the upper energy level in the subspace with fixed spin magnetization.

We start by considering a spin-1 BEC described by the following Hamiltonian~\cite{14,15}
\begin{equation}
\label{Eq1}
\hat{H} = c_2\frac{\hat{\bf L}^2}{2N}+\sum_{m_F=-1}^{1}(qm_F^2-pm_F)\hat{a}^\dagger_{m_F}\hat{a}_{m_F},
\end{equation}
under a widely used single spatial mode approximation, where a spatial wave function $\Phi({\bf r})$
is approximated to be spin independent so that the atomic field operator can be decomposed as
$\hat{\Psi}_{m_F}({\bf r})\approx \Phi({\bf r})\hat{a}_{m_F}$ with $m_F=-1,0,1$ being the magnetic spin
quantum number. Here, $N$ is the total atom number, $c_2$ is the spin-dependent interaction energy, $p\,(q)$ is linear (quadratic) Zeeman energy, and $\hat{L}_\mu = \sum_{i,j}\hat{a}^\dagger_i(F_\mu)_{ij}\hat{a}_j$ ($\mu=x,y,z$) is a total spin operator with $F_\mu$ being the spin-1 angular momentum matrix along the $\mu$ direction and
$\hat{a}_j$ ($\hat{a}_j^\dagger$) being an annihilation (creation) operator.

\begin{figure}[t]
	\includegraphics[width=3.3in]{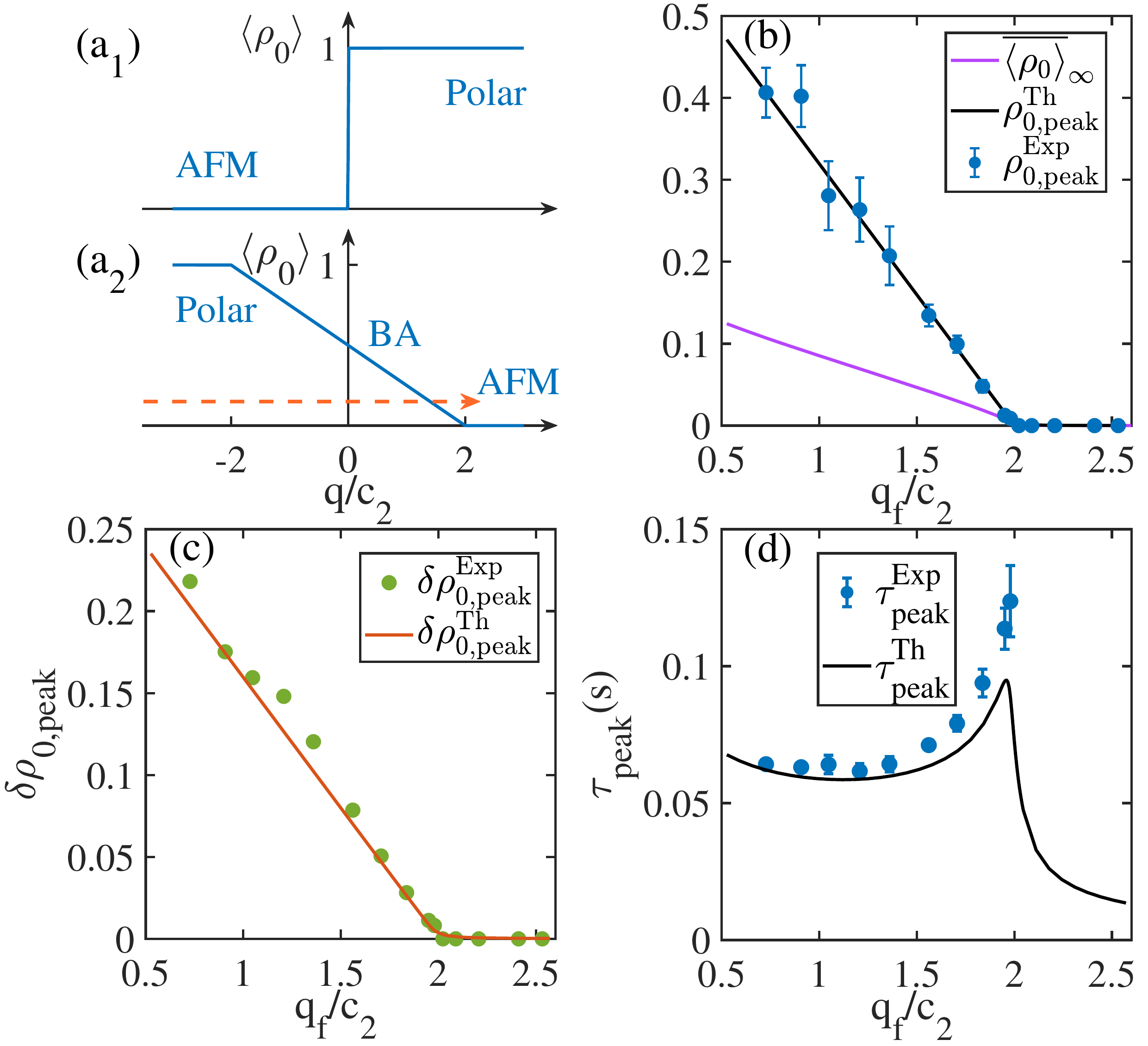}
\caption{(Color online) $\langle\rho_0 \rangle$ as a function of the quadratic Zeeman energy $q$ for ($\text{a}_\text{1}$) the ground state and
($\text{a}_\text{2}$)
the highest energy state with zero magnetization. The quench dynamics is achieved by suddenly varying $q$ from
a large negative value
to $q_f$, as schematically shown by the red arrow. Experimentally observed (b) $\rho_{0,{\textrm{peak}}}$ and
(c) $\delta \rho_{0,{\textrm{peak}}}$ with respect to $q_f$, in comparison with the theoretical results (solid lines). $\overline{\langle \rho_0\rangle}_\infty$ is plotted as a purple line.
(d) Experimentally observed occurrence time $\tau_{\textrm{peak}}$ of the first peak of $\langle \rho_0\rangle $ (solid circles), compared with the
theoretical results (black line).
Here, $c_2/h=15.2\pm0.2\,\mathrm{Hz}$.  } \label{Figure1}
\end{figure}

To explore dynamical quantum phase transitions, we prepare a condensate of sodium atoms in an AFM state
with zero magnetization [equivalent to
zero linear Zeeman energy ($p=0$)] and then suddenly change the quadratic Zeeman energy $q$ to a final value $q_f$ at $t = 0$.
As the system evolves under the final Hamiltonian, the quench dynamics can be measured.
A non-analytic change in the measured quantity as a function of the final Hamiltonian parameter $q_f$ can be
regarded as a signature of dynamical quantum phase transitions.
Since the total magnetization is conserved during the time evolution,
i.e., $[\hat{H},\hat{L}_z]=0$, the quench dynamics is restricted in the subspace with fixed eigenvalue of
$\hat{L}_z$.
For sodium atoms, which have positive $c_2$, without any linear Zeeman energy, the ground state has
a phase transition at $q=0$ from an AFM phase with equally populated atoms on the $m_F=\pm 1$ levels
to a polar phase with all atoms occupying the $m_F=0$ level [see Fig.~\ref{Figure1}($\text{a}_\text{1}$)]. After a quench, the dynamics
is restricted in the subspace with zero magnetization. In this subspace,
the highest energy level exhibits a phase transition at $q=2c_2$ between a phase with nonzero population
on the $m_F=0$ level corresponding to the BA phase in the mean-field approximation and an
AFM phase and at $q=-2c_2$ between a BA phase and
a polar phase~\cite{SM}, similar to rubidium atoms with negative $c_2$, as shown in Fig.~\ref{Figure1}($\text{a}_\text{2}$).

\begin{figure}[t]
	\includegraphics[width=3.3in]{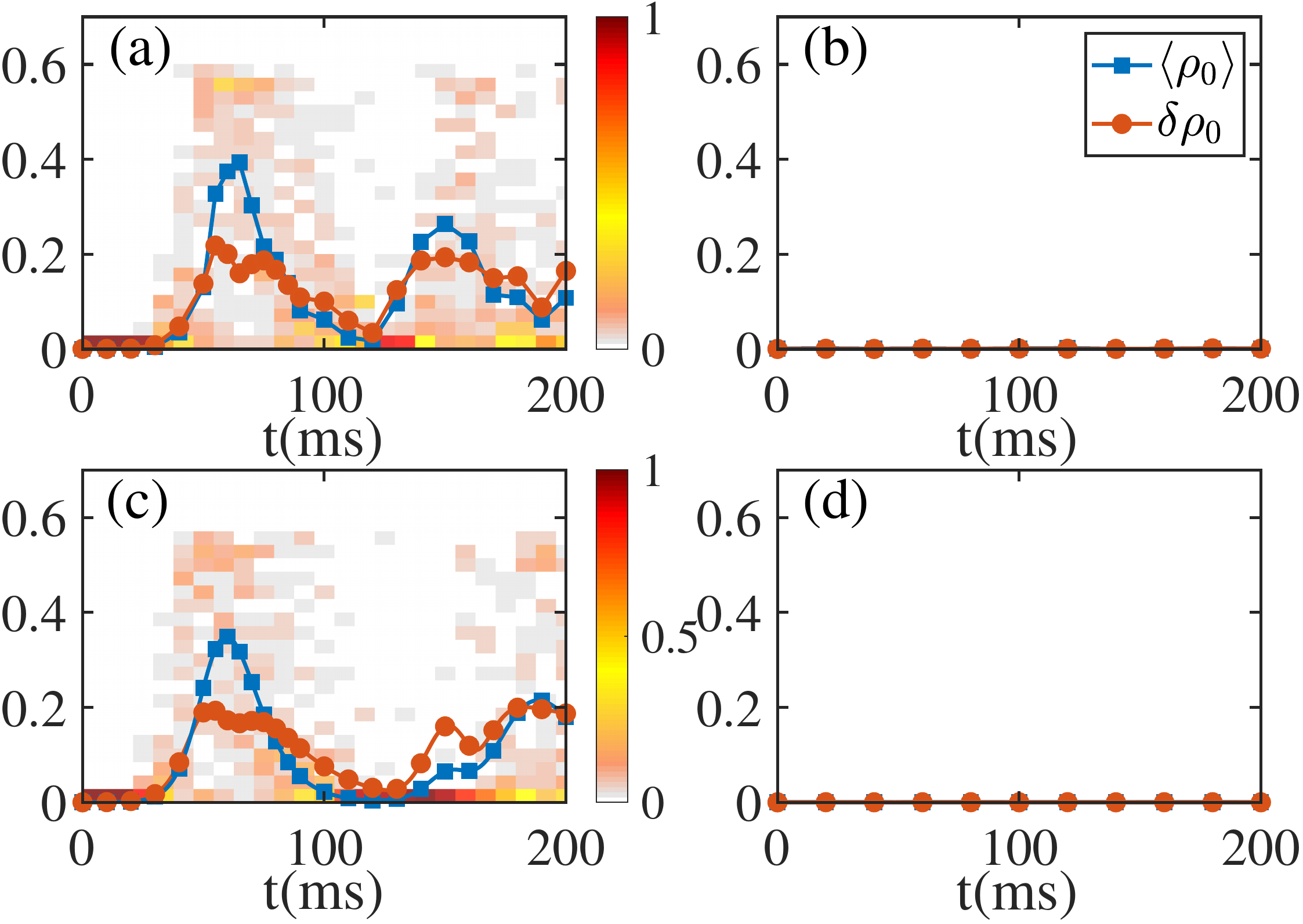}
\caption{
(Color online) Time evolution of $\langle \rho_0 \rangle$ (blue squares) and $\delta \rho_0$ (red circles)
for (a) $q_f=0.9c_2$ and (b) $q_f=2.2c_2$.
(c) and (d) plot the theoretical predicted results under the same parameters as (a) and (b), respectively.
For each $q_f$ and time, we perform $40$ times measurements.
The background squares show the probability that a measurement outcome occurs.
In (c) and (d), the probability is obtained by sampling $40$ samples using the Monte Carlo method.
Here $c_2/h=15.2\pm 0.2\,\mathrm{Hz}$ measured by spin oscillations.
} \label{Figure2}
\end{figure}

In experiments, directly detectable physical quantities are the number of atoms with spin-$m_F$ divided by
the total atom number, i.e., $\rho_{m_F} = \hat{a}^\dagger_{m_F}a_{m_F}/N$, and their average
$\langle \rho_{m_F}\rangle$ over many experimental ensembles.
A dynamical phase transition is usually characterized by an asymptotic long-time steady value of a local order parameter,
which in our case can be chosen as
$
\overline{\langle{\rho}_0\rangle}_\infty = \mathrm{lim}_{T\to\infty}\frac{1}{T}\int_0^T\langle {\rho}_0 \rangle dt
$.
Fig.~\ref{Figure1}(b) shows its increase from zero as $q_f$ is decreased from $2c_2$ (see also Ref.~\cite{Yang2019PRA}), in stark contrast to the ground state phase diagram without any phase transition at this point.
In fact, the dynamical phase transition at $q_f=2c_2$ corresponds to
the quantum phase transition of the highest energy level in the subspace with zero magnetization.
This connection can be easily explained in the mean-field approximation.
In this approximation, the ground state for $q_i=-\infty$ and the highest energy state for
$q_i>2c_2$ share the same wave function since they are both in the AFM phase with zero $\langle \rho_0\rangle$.
It follows that $\langle \rho_0\rangle$ remains zero when we suddenly vary $q$ from $-\infty$ to $q_f$ with $q_f>2c_2$.
Yet, when $q_f<2c_2$, the time evolved state is no longer an eigenstate of $\rho_0$,
leading to the appearance of nonzero values for $\langle \rho_0\rangle$ as shown in Fig.~\ref{Figure2}(a).
This picture is also valid in the many-body level given that the initial state has a significant probability to overlap with the highest energy state of the final Hamiltonian in the subspace when $q_f>2c_2$. 

In real experiments, it is a significant challenge to observe the long-time average of $\langle \rho_{m_F}\rangle$
as the long-time relaxation dynamics is unavoidable.
Fortunately, the model Hamiltonian Eq~.\eqref{Eq1} actually describes a system of N spin-1 particles with effectively infinite-range interactions~\cite{Yang2019PRA}; this enables us to characterize the dynamical phase transition by alternative
finite-time observables: $\rho_{0,\textrm{peak}}\equiv \langle \rho_0\rangle(t=\tau_{\textrm{peak}})$ and $\delta\rho_{0,\textrm{peak}}=\delta\rho_{0}(t=\tau_{\textrm{peak}})$,
the value of $\langle\rho_0\rangle$ and the standard deviation of $\rho_0$ at the first peak
of the spin oscillations, respectively [see Fig.~\ref{Figure2}]~\cite{Yang2019PRA}.
The occurrence time $\tau_\textrm{peak}$ of the first peak is around several tens of milliseconds,
making the experimental observation feasible. Indeed, the dynamical phase transition at $q_f=0$ reflecting the
ground phase transition has been experimentally demonstrated~\cite{Yang2019PRA}. However, to observe the dynamical phase
transition at $q_f=2c_2$, one needs to reduce the rapid relaxation toward the ground states for large $q_f$. We here solve
this challenging problem by significantly reducing the atom number to around $5.8\times 10^3$~\cite{SM}.

In experiments, a spin-1 BEC is produced via an all-optical procedure as detailed in Ref.~\cite{16}.
We then apply a magnetic field gradient to remove the atoms on $|m_F = \pm 1\rangle$
out of the BEC cloud~\cite{17}, followed by equilibrating the system by holding for $1\,\mathrm{s}$.
After that, we shine a $\pi/2$-pulse
radio frequency radiation to create a nearly AFM state, which has zero magnetization and
zero component on the $m_F=0$ level.
Since the experiment is very sensitive to the initial value of $\langle\rho_0\rangle$~\cite{18},
we then immediately apply
a microwave pulse for $300\,\textrm{ms}$ with a frequency of $1.7716264\,\mathrm{GHz}$,
whose detuning is zero for the clock
transition from $|F=1,m_F=0\rangle$ to $|F=2,m_F=0\rangle$
[the Rabi rate is about $1.9\textrm{kHz}$~\cite{Ketterle2003PRL} and the applied magnetic field ranges from
$0.2\,\textrm{G}$ to $0.373\,\textrm{G}$ for the experiments in Fig.~\ref{Figure1}(c)].
This pulse allows us to excite the atoms on the hyperfine level $|F=1, m_F=0\rangle$ to another level $|F=2, m_F=0\rangle$;
these atoms then escape from the trap quickly since the latter energy level
is quite unstable and the atoms on this state suffer a significant loss.
We therefore prepare the initial state with $\rho_0=0$ and $m_z=\rho_1-\rho_{-1} \simeq 0\pm 0.015$.
Note that we use a relatively weak microwave field to avoid apparent atom loss.

To study the spin dynamics, the quadratic Zeeman energy $q$ should be suddenly tuned. This can be
experimentally achieved by controlling a magnetic field or a microwave pulse, since
$q = q_M + q_B$, where $q_M$ and $q_B$ are the quadratic Zeeman energy induced by
the microwave pulse and magnetic field, respectively~\cite{19,20,21}.
During the preparation of the initial state, we fix the magnetic field so that its contribution to
the quadratic Zeeman energy is equal to our final quadratic Zeeman energy $q_f$,
i.e., $q_f = q_B \propto B^2$, which can be easily identified by measuring the Zeeman splitting induced by the magnetic field $B$. Simultaneously, we apply a resonant microwave pulse (the same pulse is also used to remove the remaining atoms on the $m_F=0$ level), generating a large negative quadratic Zeeman energy~\cite{19}.
To achieve the sudden quench, we quickly switch off the microwave pulse, leading to the final $q_f$.
After that, we perform the measurement of the fractional population $\rho_{0}$ via the standard Stern-Gerlach fluorescence imaging technique with respect to time. The experiments are repeated for 40 times at each time for each $q_f$,
and the average value $\langle{\rho}_0\rangle (t)$ and the standard deviation $\delta \rho_0 (t)$ are then determined.

In Fig.~\ref{Figure1}(b) and (c), we show our experimental results of $\rho_{0,\textrm{peak}}$ and $\delta \rho_{0,\textrm{peak}}$
as a function of $q_f$, respectively. Both quantities are zero when $q_f>2c_2$ and then exhibit a linear increase as $q_f$ decreases
when $q_f<2c_2$, which agrees well with our theoretical simulation, predicting the existence of a
second-order dynamical phase transition at $q_f=2c_2$. Fig.~\ref{Figure1}(d) further illustrates the occurrence time
$\tau_{\textrm{peak}}$ with respect to $q_f$, showing its sharp increase around $q_f=2c_2$, consistent with
the theoretical expectation that the occurrence time has a peak at $q_f=2c_2$.
Here, only the occurrence time for $q_f<2c_2$ is measured, while for $q_f>2c_2$,
the oscillation amplitude is too small to be probed. Note that for each $q_f$,
the first peak of $\langle{\rho}_0\rangle (t)$ is fitted by a Gaussian function to obtain
the occurrence time $\tau_\textrm{peak}$ and the value of $\langle{\rho}_0\rangle (t)$ at this time.
The measured dynamical phase transition corresponds to the highest energy level quantum phase transition.

\begin{figure}[t]
\includegraphics[width=2.8in]{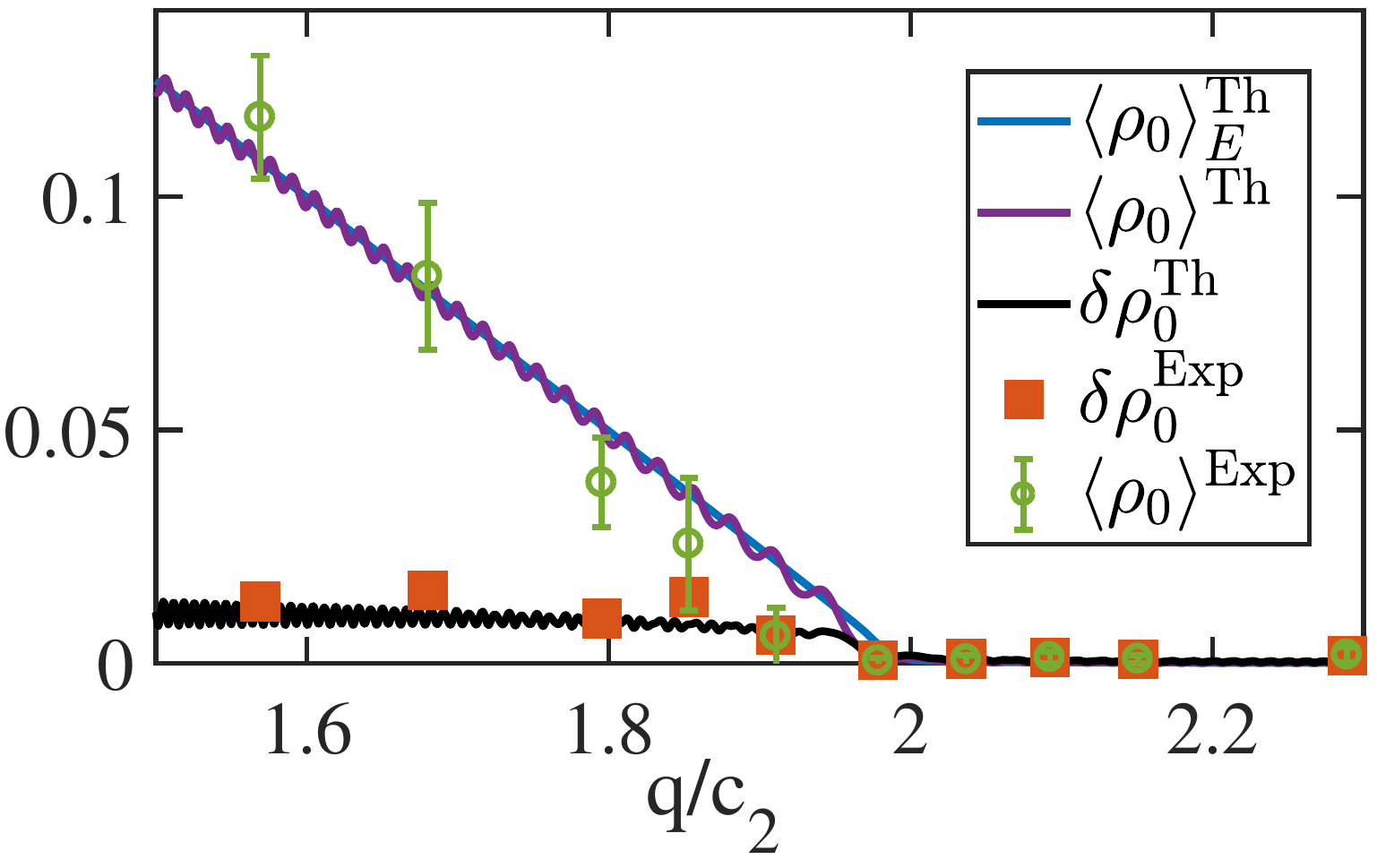}
\caption{(Color online) Quasi static measurement of the quantum phase transition in the excited state with $m_z=0$
achieved by slowly decreasing $q$ across $2c_2$ after $q$ is suddenly changed to $2.3c_2$.
The green circles and red squares denote the experimentally observed $\langle \rho_0\rangle$ and $\delta \rho_0$,
respectively, while the purple and black lines denote the numerical results of the corresponding quantities, respectively.
The blue line depicts the theoretically calculated $\langle \rho_0\rangle$ versus $q$ for the highest energy level with $m_z=0$,
while the purple line denotes this quantity for a time evolved state.
Here, $c_2/h = 13\pm 0.7\, \mathrm{Hz}$.}
\label{Figure3}
\end{figure}

Fig.~\ref{Figure2}(a) and (b) display the experimentally observed $\rho_0$ as time progresses for two typical $q_f$
across distinct phases. When $q_f>2c_2$, $\rho_0$ remains zero as time evolves
consistent with our expectation [see Fig.~\ref{Figure2}(b)and (d)]. When $q_f<2c_2$, $\rho_0$ exhibits large
fluctuations since the dynamical state is no longer an eigenstate of $\rho_0$ and each experimental measurement
gives its eigenvalue associated with a probability proportional to the occurrence times. Their average $\langle\rho_0 \rangle$ and $\delta \rho_0$ over all the ensembles exhibit an oscillation with the first peak at around $t=63\,\mathrm{ms}$. In addition, we numerically sample $\rho_0$ $40$ times via Monte Carlo sampling methods based on the theoretical probability distribution $f(\rho_0)$ of $\rho_0$ for the time evolved state.
The numerical results are plotted in Fig.~\ref{Figure2}(c) and (d), showing qualitative agreement with
the experimental results around the first peak.
However, as time further evolves, there appears the deviation that the second peak emerges earlier for the experimental
results. We attribute this deviation to the breakdown of the single spatial mode approximation~\cite{Yang2019PRA}. Since the time evolved state after the quench corresponds to the higher energy levels of the final Hamiltonian for the spin degrees of freedom for zero magnetization, the atoms can relax their energy stored in the spin degrees of freedom into the spatial degrees of freedom,
resulting in the spatial mode excitation so that atoms do not share the same spatial wave function, breaking down the
single mode approximation. In fact, such a relaxation process is strongly enhanced for larger $c_2$ probably due to inelastic collisions, hindering the observation of the dynamical phase transition~\cite{SM}.

\begin{figure}[t]
	\includegraphics[width=3.3in]{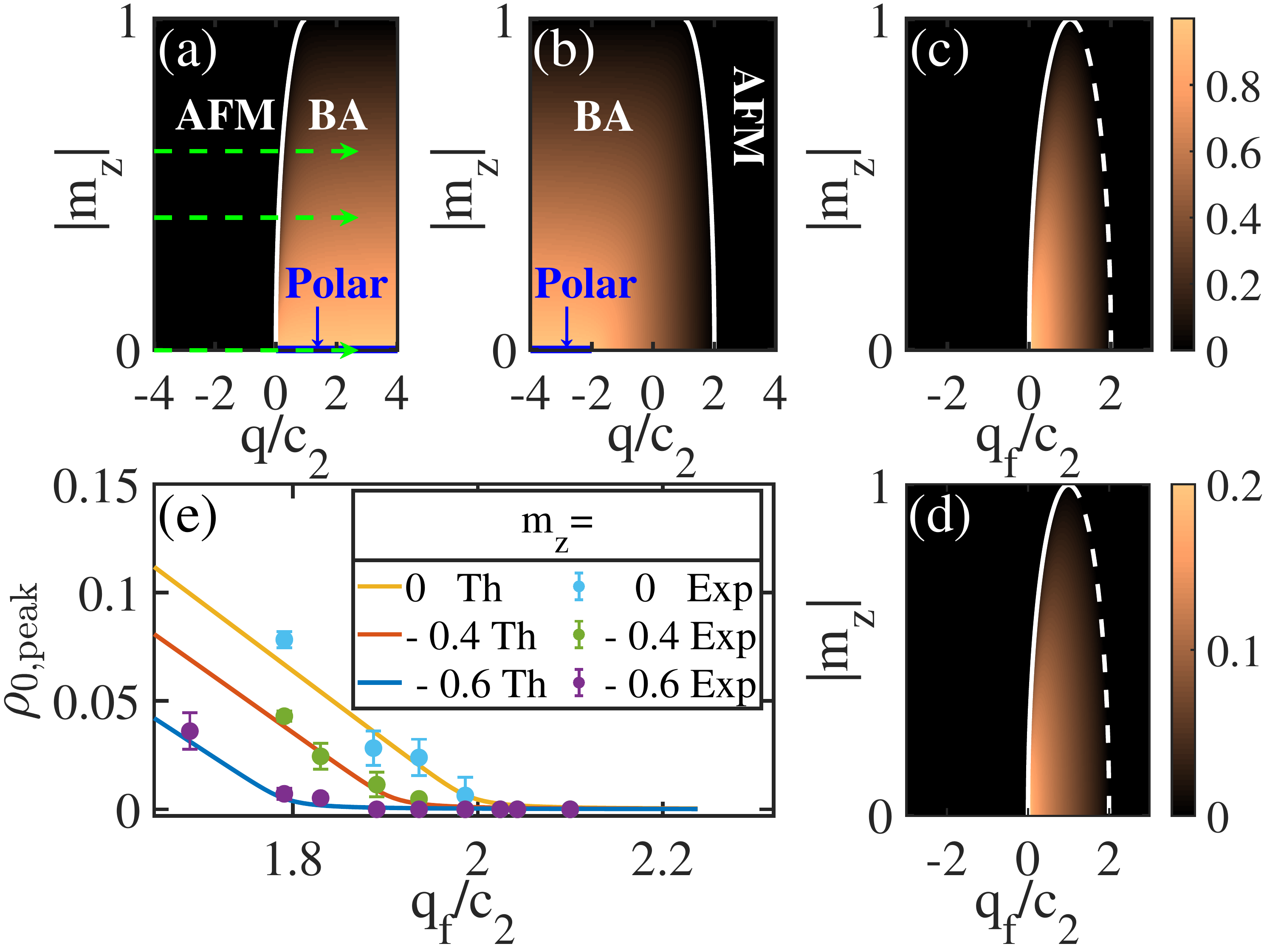}
 \caption{(Color online) Theoretically calculated $\langle \rho_0\rangle$ with respect to $q$ and $m_z$ for (a) the ground state and (b) the highest energy level
 for a fixed $m_z$. With nonzero $m_z$, there are two distinct phases separated by white lines. The solid bold blue lines
 show the polar phase with $\rho_0=1$ for zero $m_z$.
 Theoretically calculated (c) $\rho_{0,\textrm{peak}}$ and (d) $\overline{\langle\rho_0 \rangle }_\infty$
 in the $q_f$ and $|m_z|$ plane. (e) Experimentally observed $\rho_{0,\textrm{peak}}$ (solid circles) for $m_z=0,-0.4,-0.6$ [see arrows in (a)], in comparison with the theoretical results (solid lines).
 Here, $c_2/h=14.3\pm 0.5\,\mathrm{Hz}$.
 } \label{Figure4}
\end{figure}

To show the presence of the quantum phase transition in the excited state with $m_z=0$, we have further performed the quasi static measurement of the phase transition in the excited state.
This is experimentally achieved by quickly varying $q$ from a large negative value to $q_f=2.3c_2$ followed
by slowly tuning $q$ across the transition point by $q=q_f-vt$ with $v=3\,\textrm{Hz/s}$.
As time evolves, we perform the measurement of $\rho_0$. In Fig.~\ref{Figure3}, we plot the measured $\langle \rho_0\rangle$
and $\delta \rho_0$, which are in qualitative agreement with the numerical simulation results.
The figure also demonstrates that even for the numerical simulation (see the purple solid line), the transition point
is slightly smaller than $2c_2$. This arises from the closing of the energy gap between the highest energy state and its neighboring
energy level, leading to
an impulse region where the state remains unchanged so that $\rho_0$ cannot adapt to the system change instantaneously.
To achieve the precise identification of the transition point, we need to
control $q$ to vary very slowly. However, such a slow variation takes a
long time, inevitably involving the energy transfer into the spatial modes. Therefore, the quench dynamics provides an ideal method to identify the excited state quantum phase transition.

We now study the dynamical phase transition for finite spin magnetization $m_z$. In Fig.~\ref{Figure4}(a-b), we
map out the ground state and the highest energy level (in a subspace with fixed $m_z$) phase diagram in the $(q,|m_z|)$ plane, respectively. When $m_z\neq 0$,
both of these two levels exhibit two distinct phases: the AFM phase with $\rho_0=0 $
and the BA phase with nonzero $\langle\rho_0\rangle$.
As $|m_z|$ rises from 0,
the critical points for the former slightly increase from 0 [see the white line
in Fig.~\ref{Figure4}(a)] and for the latter slightly decrease from $2c_2$ [see the white line
in Fig.~\ref{Figure4}(b)].
For the former (latter), the left (right) region corresponds to the AFM phase while the right (left) one to the BA phase.
Starting with a state corresponding to an AFM phase for a large negative quadratic Zeeman energy $q$,
we suddenly tune $q$ to $q_f$ and then calculate
$\rho_{0,\textrm{peak}}$ and $\overline{\langle \rho_0\rangle}_\infty$ as time evolves.
Fig.~\ref{Figure4}(c) and (d) plot these two quantities in the
plane $(q_f,|m_z|)$, respectively, illustrating dynamical phase transitions for positive $q_f$, the boundary of which
is related to the phase transition boundary of the highest energy level for a fixed $m_z$
(described by the dashed white lines).

In experiments, we prepare the BEC in an AFM state as previously described.
We then apply a microwave pulse for $10\,\mathrm{ms}$
to excite atoms from the hyperfine level $|F=1, m_F=+1\rangle$ to $|F=2, m_F=0\rangle$.
Since the lifetime of the atoms on the level $|F=2, m_F=0\rangle$ is very short, this operation
decreases the number of atoms on $|F=1, m_F=+1\rangle$. Using this procedure, we are able to prepare a state
with different $m_z$ by tuning the microwave frequency.
After that, we immediately apply a microwave pulse for $290\,\mathrm{ms}$ to pump atoms on $|F=1, m_F=0\rangle$ to $|F=2, m_F=0\rangle$;
this process removes all atoms on $|F=1, m_F=0\rangle$ for a fixed $m_z$ while keeping the quadratic
Zeeman energy a large negative value. Finally, we suddenly switch off the microwave radiation, leading to
a sudden change of the quadratic
Zeeman energy, and then perform a measurement for $\rho_0$ as time evolves. Our experimental results for three distinct $m_z$ are shown in Fig.~\ref{Figure4}(e). We see clearly the decrease of the critical phase transition points as $|m_z|$ increases, which agrees well with theoretical prediction.

In summary, we have experimentally studied the dynamical phase transition in a spinor condensate
by suddenly tuning the quadratic Zeeman energy. The dynamical phase transition is demonstrated by the appearance of a
non-analytical change in the spinor atom number as a function of a final Hamiltonian parameter.
We find that the dynamical phase transition has a correspondence with the highest energy level phase transition
for both cases of zero and finite magnetization.


\begin{acknowledgments}
We thank Yingmei Liu, Ceren Da\u{g}, and Anjun Chu for helpful discussions. This work was supported by the Frontier Science Center for Quantum Information of the Ministry of Education of China, Tsinghua University Initiative Scientific Research Program, and the National key Research and Development Program of China (2016YFA0301902). Y. X. also acknowledges the support
by the start-up fund from Tsinghua University,
the National Thousand-Young-Talents Program and the National Natural Science Foundation
of China (11974201).
\end{acknowledgments}

\begin{widetext}
\section{Supplemental Material}

\setcounter{equation}{0} \setcounter{figure}{0} \setcounter{table}{0} %
\renewcommand{\theequation}{S\arabic{equation}} \renewcommand{\thefigure}{S%
\arabic{figure}} \renewcommand{\bibnumfmt}[1]{[S#1]} \renewcommand{%
\citenumfont}[1]{S#1}

In the supplementary material, we will show the presence of singularities in the energy of the highest excited state in a
subspace with zero magnetization and show the effects of $c_2$ on the relaxation process.

To illustrate the existence of a singularity in the energy of the highest excited state, we plot the level's energy in Fig.~\ref{figs1}. Clearly, the second derivative of the energy
with respect to $q$ exhibits a discontinuous jump at $q=\pm 2 c_2$, implying the existence of a second-order excited state
quantum phase transition there. This is consistent with the existence of a discontinuous jump for the first derivative of the order parameter $\langle \rho_0 \rangle$ with respect to $q$ [see Fig. 1($\text{a}_\text{2}$)].
	
To show the effects of $c_2$ on the relaxation process, in Fig.~\ref{figs2}, we plot the measured $\langle \rho_0\rangle $ as a function of time after $q$ is suddenly quenched to $q_f=2.1c_2$
for different $c_2$, which is controlled by tuning the atom number $N$, given $c_2\propto N^{2/5}$ under Thomas-Fermi approximation.
The figure demonstrates that while $\langle \rho_0\rangle$
remains smaller than $0.4\%$ for small $c_2$ (there are no observable atoms for $\rho_0$ except for the noise of a camera), it increases from zero for sufficiently large values of $c_2$, implying that the system decays
toward the ground state of the final Hamiltonian with $\langle{\rho}_0\rangle = 1$.
Our results are
consistent with previous observation that the relaxation is stronger for larger $q_f$ and $c_2$~\cite{Liu2009PRL,Yang2019PRA}.

\begin{figure}[t]
\includegraphics[width=6.4in]{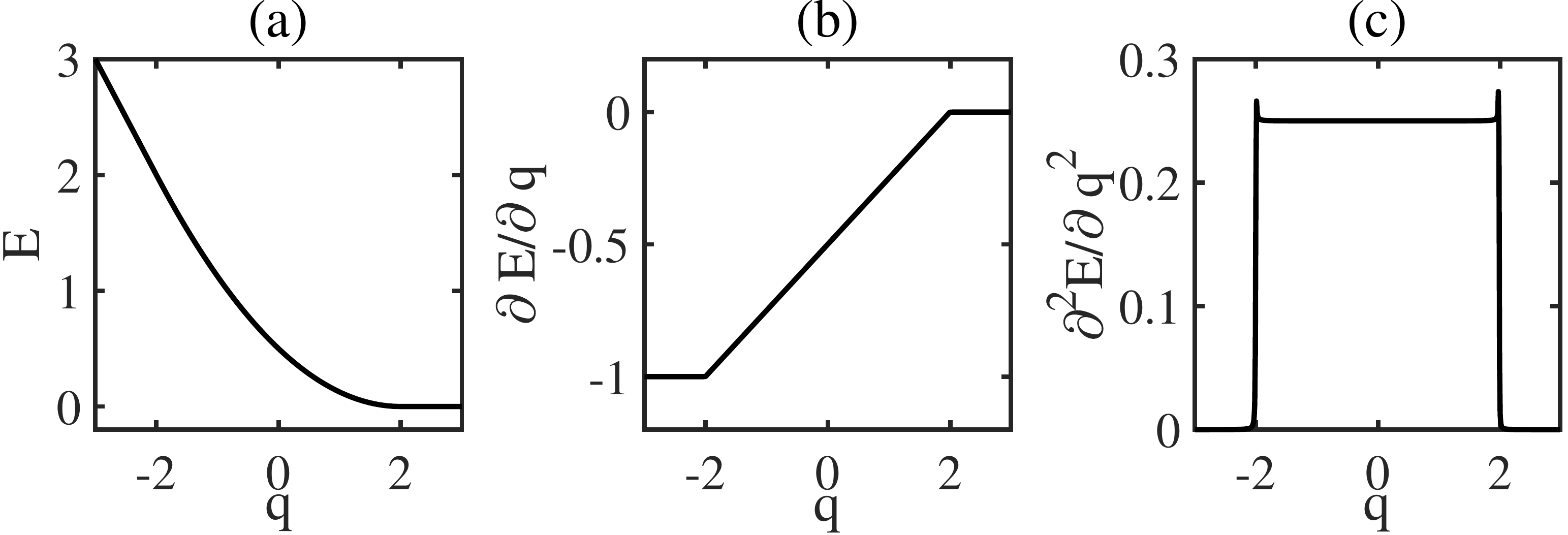}
\caption{(Color online) Theoretically calculated (a) energy per particle of the highest excited state in a subspace with zero
magnetization, (b) its first and (c) second derivative with respect to $q$. The units of $E$ and $q$ are $c_2$.}
\label{figs1}
\end{figure}

\begin{figure}[t]
	\includegraphics[width=2.6in]{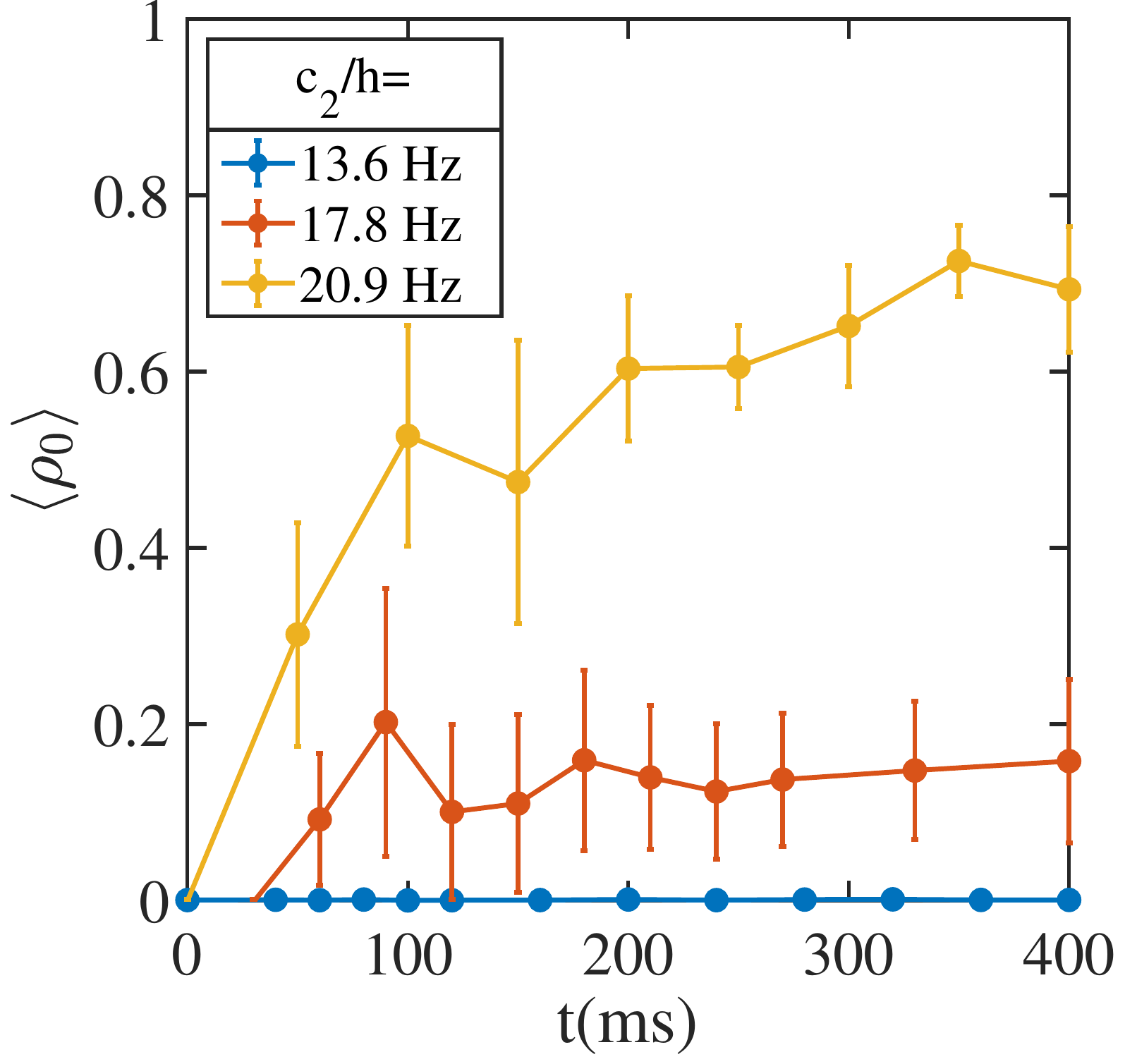}
 \caption{(Color online) Experimentally measured $\langle \rho_0\rangle$ as a function of time for distinct $c_2$. As
 the interaction strength $c_2$ is increased by raising the atom number, $\langle \rho_0\rangle$ develops nonzero values instead of remaining zero as time progresses, reflecting that the atoms tend to decay into the ground state with $\rho_0=1$ of the final Hamiltonian. Here, $q_f\approx2.1c_2$.
  } \label{figs2}
\end{figure}

\end{widetext}

\end{document}